\documentclass[%
 reprint,
showpacs,preprintnumbers,
bibnotes,
 amsmath,amssymb,
 aps,
]{revtex4-1}

\usepackage{graphicx}
\usepackage[english]{babel}
\usepackage{microtype}          
\usepackage[breaklinks=true,colorlinks=true,linkcolor=blue,urlcolor=blue,citecolor=blue]{hyperref}

\usepackage{xcolor}     
\usepackage{mathtools}
\usepackage{braket}
\usepackage{tikz}

 \usepackage{longtable}
  \usepackage{booktabs}
 \usepackage{siunitx}
 \usepackage{csquotes}

\newcommand{\lowerbossphantom}{\vphantom{\bar{\bar{x}}}}
\newcommand{\upperbossphantom}{\vphantom{\dagger}}
\newcommand{\tempop}[3][\textstyle]{\settowidth{\dimen1}{$#1\hat{#2}$}\makebox[\dimen1][l]{$#1\hat{#2\mspace{#3}}$}}
\newcommand{\xop}[1]{{\mathchoice{\tempop[\displaystyle]{#1}{3.5mu}}{\tempop{#1}{3.5mu}}{\tempop[\scriptstyle]{#1}{3.5mu}}{\tempop[\scriptscriptstyle]{#1}{3mu}}}}
\newcommand{\chat}[1]{\ensuremath{\xop{#1}}}
\newcommand{\aop}[2]{\ensuremath{\chat{c}_{#1#2\lowerbossphantom}^{\upperbossphantom}}}
\newcommand{\cop}[2]{\ensuremath{\chat{c}_{#1#2\lowerbossphantom}^{\dagger\upperbossphantom}}}

\usepackage{dcolumn}
\usepackage{bm}

\begin{document}
\bibliographystyle{apsrev}
\preprint{APS/123-QED}

\title{Doublon formation by ions impacting a strongly correlated finite lattice system}

\author{Karsten Balzer$^1$,
Maximilian Rodriguez Rasmussen$^{2}$, Niclas Schl\"unzen$^{2}$, Jan-Philip Joost$^{2}$, and
Michael Bonitz$^{2}$}
 \email{bonitz@theo-physik.uni-kiel.de}
\affiliation{
$^1$Rechenzentrum,
Christian-Albrechts-Universit\"{a}t zu Kiel, D-24098 Kiel, Germany\\
$^2$Institut f\"ur Theoretische Physik und Astrophysik, 
Christian-Albrechts-Universit\"{a}t zu Kiel, D-24098 Kiel, Germany
}

\date{\today}

\begin{abstract}
Strongly correlated systems of fermions have a number of exciting collective properties. Among them, the creation of a lattice that is occupied by doublons, i.e. two quantum particles with opposite spins, offers interesting electronic properties. In the past a variety of methods have been proposed to control doublon formation, both, spatially and temporally. Here, a novel mechanism is proposed and verified by exact diagonalization and nonequilibrium Green functions simulations---fermionic doublon creation by the impact of energetic ions. We report the formation of a nonequilibrium steady state with homogeneous doublon distribution. 
The effect should be observable in strongly correlated solids in contact with a high-pressure plasma and in fermionic atoms in optical lattices.
\end{abstract}

\maketitle
%

Strongly correlated systems are attracting increasing interest in many fields including dense plasmas \cite{fortov_07}, warm dense matter \cite{dornheim_prl}, dusty plasmas \cite{bonitz_rpp_10} and ultracold atoms \cite{schneider_np_12}. Among the most intriguing phenomena in strongly correlated quantum systems of both, fermions and bosons, is the formation of doublons---pairs of repulsively bound particles occupying the same lattice site \cite{winkler_nat_06}. 
In recent years, there have been many attempts to study the dynamics of doublons after a correlated system is driven out of equilibrium leading to many surprising results. ``Quantum distillation''---the spatial separation of doublons and single fermions---was observed in Ref.~\cite{heidrich_pra_09, natphys15}. The nonequilibrium expansion dynamics of a fermionic particle cloud following a confinement quench and its slowing down due to doublon formation has been studied experimentally in a 2D optical lattice \cite{schneider_np_12} and theoretically by 2D quantum simulations using nonequilibrium Green functions (NEGF) \cite{schluenzen_prb_16}.
Also, the external control of doublons 
by an interaction quench \cite{schecter_pra_12},  by external
electric fields \cite{eckstein_prl_10,eckstein_jpcs_13,balzer_epl_14,genske_14,joura15,kolovsky16} or by optical excitation \cite{ligges_17} has been proposed. Furthermore, the dynamics of heteronuclear doublons \cite{covey_ncom_16} and the spatial transfer of doublons via topological edge states \cite{bello_prb_17} have been studied.

Previous setups of doublon manipulation mostly involved spatially homogeneous systems containing a large number of fermions triggering their collective response to a spatially delocalized excitation.
In contrast, in this Letter we predict a novel mechanism to induce and control the formation of doublons in a \textit{finite system} where the \textit{excitation is localized in space and time} and driven by energetic ions penetrating a strongly correlated material and depositing energy (``stopping power'', e.g. \cite{zhao, puska14, balzer_prb_16}). The mechanism is demonstrated by exact diagonalization simulations, and a physical explanation is given with an analytical model in terms of the Landau--Zener effect \cite{landauzener32}. We then extend the analysis to large isolated Hubbard clusters in one and two dimensions containing up to $54$ sites. By performing extensive, long NEGF simulations we demonstrate the emergence of a stationary nonequilibrium state with homogeneous doublon distribution.
Finally, we demonstrate how the effect can be further enhanced by using a sequence of excitations. 

\begin{figure}[t]
\includegraphics[width=0.48\textwidth]{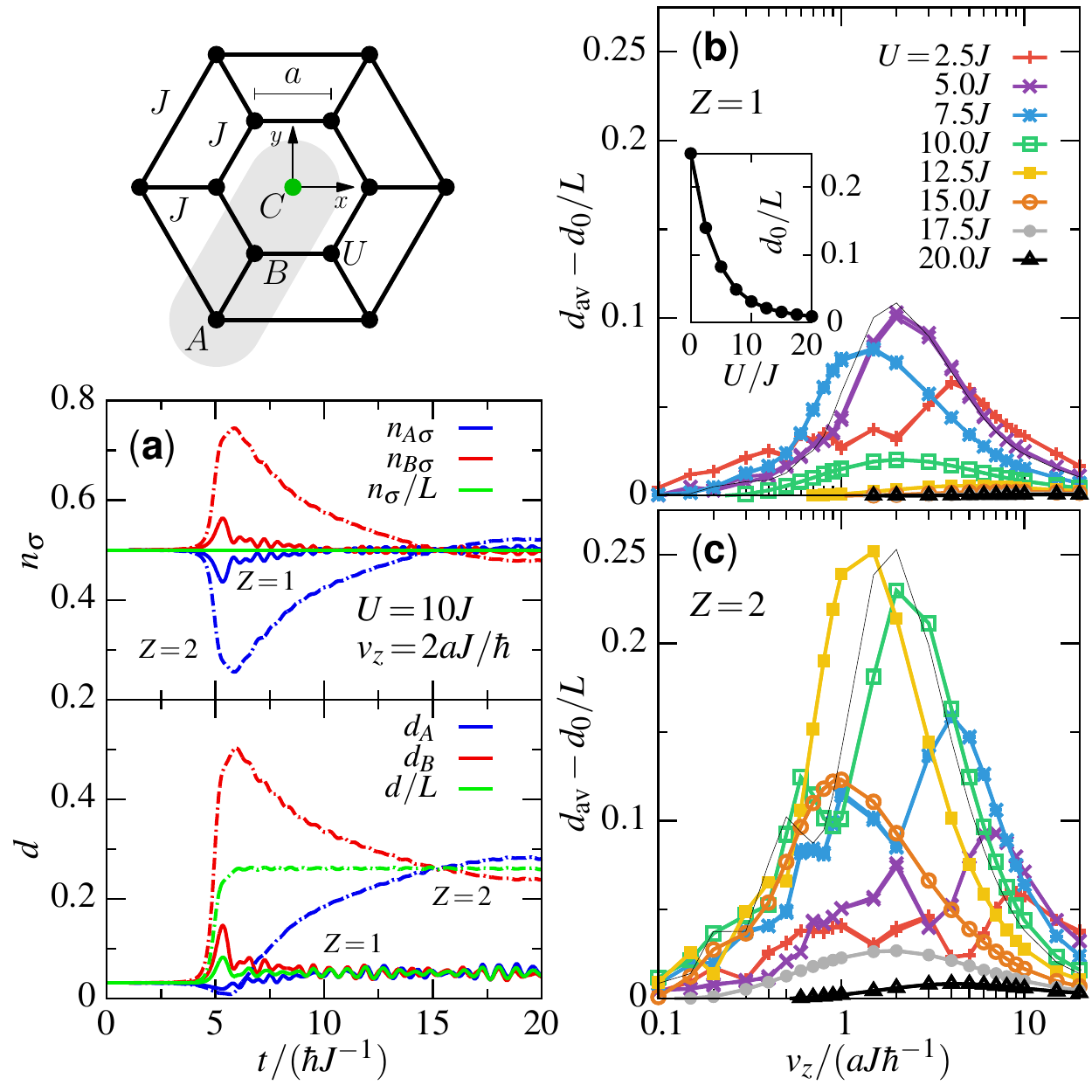}
\caption{\label{fig.honeycomb}Ion impact-induced doublon formation in a two-dimensional Hubbard nano-cluster (top left, black points) with $L=12$ sites, nearest-neighbor hopping $J$ and on-site interaction $U$. \textbf{a)} Time evolution of the electron density $n_{i\sigma}$ (top) and double occupation $d_i$ (bottom) on sites $A$ (blue) and $B$ (red) for a positive charge with $Z=1$ (solid lines) and $Z=2$ (dash-dotted lines) impacting the system at $U=10J$ with velocity $v_z=2aJ/\hbar$ in point $C=(0,0,0)$. The green curves show the overall density and double occupation, $n_\sigma(t)/L=\tfrac{1}{L}\sum_i n_{i\sigma}(t)=0.5$ and $d(t)/L=\tfrac{1}{L}\sum_i d_i(t)$, respectively. \textbf{b) and~c)}:~Change of the double occupation, $d_\textup{av}-d_0/L$, as function of $v_z$ for different $U$ and $Z=1$ and $Z=2$, respectively. The thin grey curves correspond to $U=5.4J$ and $10.8J$, respectively.}
\end{figure}

\textbf{Model.} We consider strongly correlated electrons in a single-band Hubbard model with nearest-neighbor hopping $J$ and on-site interaction $U$,
\begin{align}
\label{eq.hm}
\chat{H}=-J\sum_{\langle i,j\rangle\sigma}\cop{i}{\sigma} \aop{j}{\sigma}+U\sum_{i}\chat{n}_{i\uparrow}\chat{n}_{i\downarrow}+\sum_{i\sigma} W_i(t) \chat{n}_{i\sigma}\,,
\end{align}
where $\chat{n}_{i\sigma}=\chat{c}_{i\sigma}^\dagger \chat{c}_{i\sigma}$ is the density and $\sigma$ denotes the spin. The third term describes the interaction of the electrons with a moving charge (ion, projectile), where the time dependence of $W_i(t)$ is parameterized by a classical trajectory $\bm{r}(t)$. For an ion of positive charge $Ze$ and lattice sites located at $\bm{r}_i$, we use $W_i(t)=-Ze^2/(4\pi\epsilon_0 |\bm{r}(t)-\bm{r}_i|)$, neglect nonlocal contributions in the electron--ion interaction ($W_{ij}=\delta_{ij}W_i$), and measure energies, times and lengths in units of $J$, $\hbar J^{-1}$ and the lattice constant $a$, respectively. Furthermore, we define $W_0=e^2/(4\pi\epsilon_0 a)$ and set $W_0=14.4J$, which corresponds to a force $J/a=1$\,eV/\AA.

In Fig.~\ref{fig.honeycomb}, we present solutions of the system (\ref{eq.hm}) by time-dependent exact diagonalization for a 2D half-filled Hubbard nano-cluster with $L=12$ sites, starting at $t=0$ from the ground state. The trajectory of the ion is set to $\bm{r}(t)=(0,0,z+v_zt)$ with velocity $v_z$ and initial $z$-position such that $W_i(t=0)\rightarrow 0,\;\forall i$. Figure~\ref{fig.honeycomb}a shows the time evolution of the electron density and double occupation,
\begin{align}
    n_{i\sigma}(t)&=\langle \chat{n}_{i\sigma}\rangle(t)\,,&
    d_i(t)&=\langle \chat{n}_{i\uparrow}\chat{n}_{i\downarrow}\rangle(t)\,,
\end{align}
for an on-site interaction $U=10J$, $v_z=2aJ/\hbar$ and $Z=1$ and $2$, where the expectation values are computed as $\langle\chat{\cal{O}}\rangle(t)=\langle\psi(t)|\chat{\cal{O}}|\psi(t)\rangle$, with the many-electron wave function $|\psi(t)\rangle=\exp\{-\tfrac{\textup{i}}{\hbar}\!\int_{0}^t\!ds\chat{H}(s)\}|\psi(0)\rangle$. During the time of impact ($t=5\hbar J^{-1}$), both $n_{B\sigma}$ and $d_B$ ($n_{A\sigma}$ and $d_A$) increase (decrease). After departure of the projectile the electron densities return (close to) their initial value $n_{i\sigma}=0.5$. In contrast, the spatiotemporal evolution of the double occupation~\cite{hofmann_12} is such that $d_{A,B}$ remain at a higher value than the initial value, particularly for $Z=2$. Thus, the interaction with the projectile has created a significant number of doublons, indicating the emergence of a stationary nonequilibrium (``pre-thermalized'' \cite{moeckel_08,eckstein_prl_08}) state. 
For a quantitative analysis, we follow 
the site- and time-averaged double occupation $d_\textup{av}=\tfrac{1}{\Delta t L}\int_{\Delta t}\!dt\sum_i\langle \chat{n}_{i\uparrow}\chat{n}_{i\downarrow}\rangle$, which is shown in Figs.~\ref{fig.honeycomb}b and \ref{fig.honeycomb}c. A striking result is the non-monotonic dependence of $d_\textup{av}$ on the projectile velocity with a maximum around $v_z \sim (1\dots 3) aJ/\hbar$. Moreover, also the dependence on $U$ is non-monotonic: $d_\textup{av}$ exhibits a single maximum which is in the range of $U\sim 5 J$, for $Z=1$, and $U\sim 12$, for $Z=2$. Further,  $d_\textup{av}$, increases with the projectile charge. We note that in the present setup we consider a projectile with constant kinetic energy, for a discussion on the energy transfer see~\cite{balzer_prb_16}.

{\bf Analytical model}. To understand the main mechanism of the doublon formation, we consider a Hubbard dimer at half-filling and develop a Landau--Zener (LZ) description~\cite{landauzener32}. The dimer is excited by a time-dependent energy $W(t)=-W_0\exp[-t^2/(2\tau^2)]$ on one site, which well mimicks the projectile, where $\tau>0$ is linked to the inverse projectile velocity (see below), and we use $W_0=2U$. In the basis $\{|\!\uparrow,\downarrow\rangle,|\!\downarrow,\uparrow\rangle,|\!\uparrow\downarrow,0\rangle,|0,\uparrow\downarrow\rangle\}$, the dimer hamiltonian reads
\begin{align}
\label{eq.hm.dimer}
\chat{H}_\textup{dimer}(t)=\left(
\begin{array}{cccc}
 W(t) & 0 &   -J & -J\\
 0 & W(t) &    J &  J\\
-J & J & U+2W(t) &  0\\
-J & J &    0 & U
\end{array}
\right)\,,
\end{align}
and is straightforwardly diagonalized for all times.

\begin{figure}[t]
\includegraphics[width=0.48\textwidth]{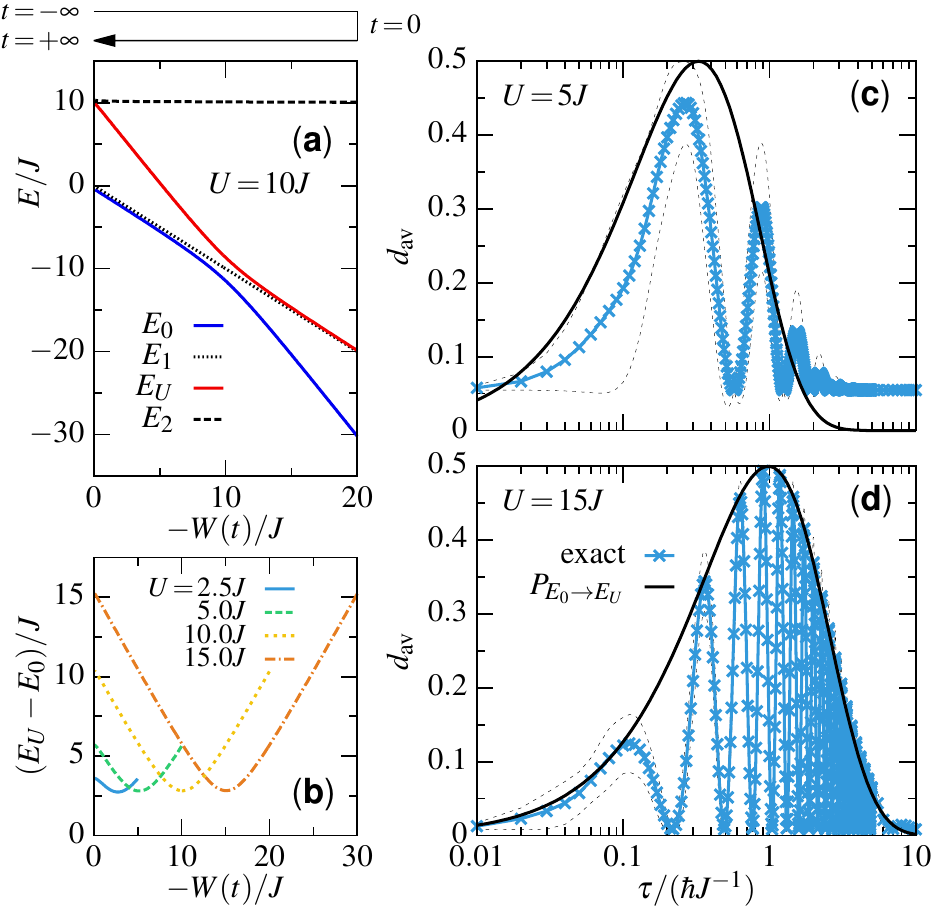}
\caption{\label{fig.dimer1} Hubbard dimer of Eq.~(\ref{eq.hm.dimer}). \textbf{a)}~Evolution of the eigenenergies as function of $W(t)$, for $U=10J$. The initial and final state correspond to $W=0$, and the impact of the projectile to $-W/J=20$, its trajectory is sketched above the figure. \textbf{b)} Eigenenergy difference, $E_U-E_0$, as function of $W(t)$ for  $W_0=2U$ and different values of $U$. \textbf{c) and~d)}: Average double occupation $d_\textup{av}$ (see text) in the dimer (blue lines) for $t\gg\tau$, $U=5J$ and $15J$ as function of $\tau$, averaged over a time window of $\Delta t=50\hbar J^{-1}$ (the thin dashed lines mark the minimum and maximum values of the double occupation). The black line represents the Landau--Zener result based on Eq.~(\ref{eq.hm.dimer.lz}). }
\end{figure}

Figure~\ref{fig.dimer1}a shows the evolution of all four eigenenergies for $U=10J$ as function of $W(t)$. Starting in the triplet ground state ($E_0$) for $t\ll-\infty$, the dimer undergoes a transition to the second excited state ($E_U$) via an avoided crossing when $W(t)$ is switched on sufficiently fast. Using a reduced two-level Landau--Zener picture, the probability to find the dimer after the full excitation, i.e., for $t\gg+\infty$, in the state $E_U$ can be approximated by a twofold (back-and-forth) passage of the avoided level crossing:
\begin{align}
\label{eq.hm.dimer.lz}
P_{E_0\rightarrow E_U}=2p(1-p)\,,
\end{align}
where $p$ denotes the Landau--Zener 
transition probability for a single diabatic passage of the crossing,
\begin{align}
\label{eq.hm.dimer.lz-single1}
p&=\exp\left(-\frac{2\pi V^2}{\hbar|dE/dt|}\right)\,,
\quad V=\frac{1}{2}\text{min}_{W(t)}\,E\,,
\end{align}
and $E=E_U-E_0$. To evaluate Eqs.~(\ref{eq.hm.dimer.lz-single1}) and (\ref{eq.hm.dimer.lz}), we use $\frac{dE}{dt}=\frac{dE}{dW}\frac{dW}{dt}=-\frac{dE}{dW}\frac{tW(t)}{\tau^2}$, set $t=\pm\tau$ [turning points of $W(t)$] and obtain 
\begin{align}
\label{eq.hm.dimer.lz-single2}
p=\exp\left(-\frac{2\pi V^2\exp(1/2)\tau}{\hbar W_0 |dE/dW|}\right)\,.
\end{align}
From Fig.~\ref{fig.dimer1}b, we furthermore observe that $V$ and $dE/dW$ are almost independent of $U$: $2V\approx2.826J$ and $|dE/dW|\approx0.976$, around $W(t)=-W_0$, therefore, the probabilities $p$ and $P_{E_0\rightarrow E_U}$ only depend (for fixed $U$) on the temporal width $\tau$ of the excitation. 

Figures~\ref{fig.dimer1}c and \ref{fig.dimer1}d show the average double occupation, $d_\textup{av}$, in the dimer, from the Landau--Zener model, for $U=5J$ and $15J$ (black curves), together with the exact dynamics of Eq.~(\ref{eq.hm.dimer}). For sufficiently large $U\gtrsim10J$, we find that our model (\ref{eq.hm.dimer.lz}) reproduces the envelope of $d_\textup{av}$ very well (a doublon is created on site one with maximum probability $1/2$) although it does not capture the oscillations as a function of $\tau$ that are proportional to the field $W_0$ and are due to superimposed transient Bloch oscillations~\cite{eckstein_11}.

With insight from the dimer model, we find parameters for particularly efficient doublon formation in Fig.~\ref{fig.honeycomb}:~(i)~the optimal on-site interaction is $U^*\approx\tfrac{1}{2}W^*$, where $W^*=Z\cdot 10.8J$ denotes the maximum induced potential averaged over the sites $A$ and $B$; thus $U^*=5.4J$ for $Z=1$ and $U^*=10.8J$ for $Z=2$, cf. the thin solid lines in Figs.~\ref{fig.honeycomb}b and \ref{fig.honeycomb}c. (ii)~For $U=U^*$, the velocity $v_z^*$, that maximizes the doublon yield, decreases linearly with the ion charge $Z$. This follows from the Landau--Zener condition $\tfrac{d}{d\tau}P_{E_0\rightarrow E_U}(\tau)=0$ which is solved by $\tau=\tau^*$:
\begin{align}
(\tau^*)^{-1}=\frac{2\pi V^2\exp(1/2)}{\hbar W_0|dE/dt|\log(2)}\propto v_z^*\,.
\end{align}

\begin{figure}[h]
\includegraphics[width=0.48\textwidth]{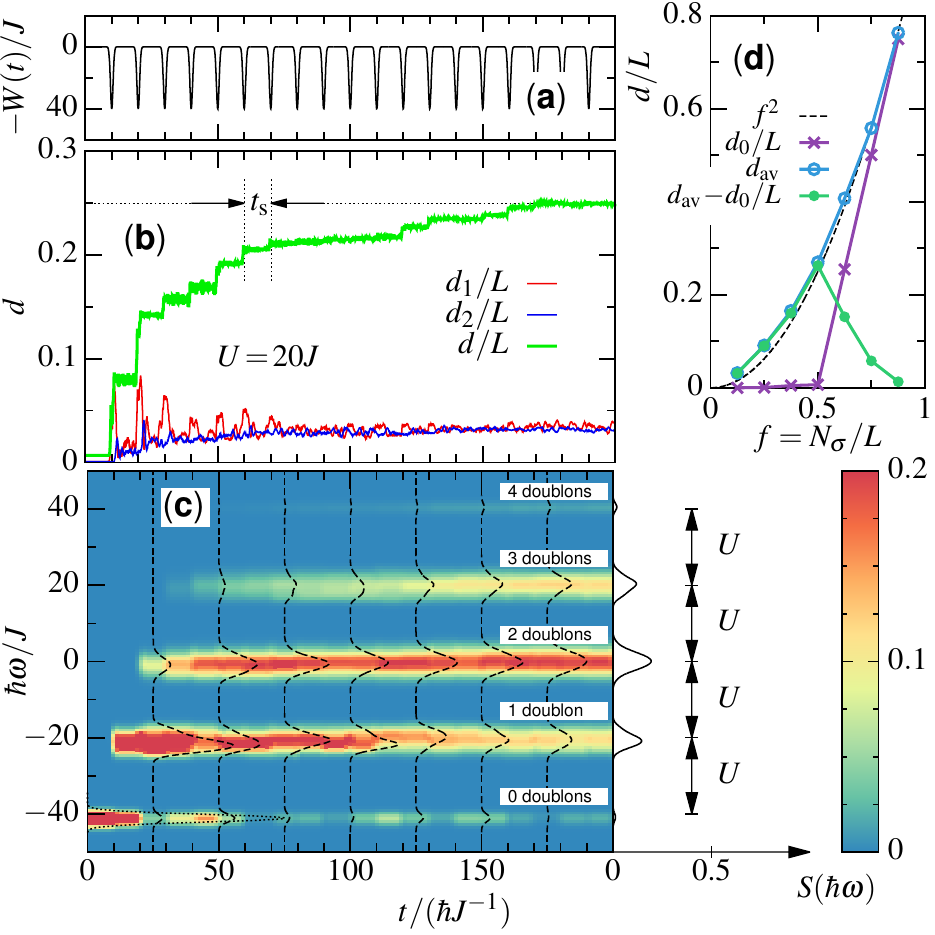}
\caption{\label{fig.dimer2} Periodic excitation of a half-filled Hubbard chain with $L=8$ sites and $U=20J$. \textbf{a)}~Applied field, $W(t)$, with $\tau=0.5\hbar J^{-1}$ and a temporal separation $t_\textup{s}=10\hbar J^{-1}$ of the peaks. \textbf{b)}~Time evolution of the mean double occupation $d/L=\tfrac{1}{L}\sum_id_i(t)$. \textbf{c)}~Time evolution of the energy spectrum $S(\hbar\omega,t)$, Eq.~(\ref{eq.hm.chain.spectrum}) with $\hbar\omega_0=J$. \textbf{d)}~Doublon formation process for different filling fractions $f=N_\sigma/L$, where $N_\sigma$ denotes the number of electrons of spin $\sigma$.}
\end{figure}

{\bf Multiple excitations and larger systems}. A verification that the Landau--Zener process drives the formation of doublons also in larger systems is presented in Fig.~\ref{fig.dimer2} for a 1D half-filled Hubbard chain with $L=8$ sites, $U=20J$ and periodic boundary conditions. 
Keeping the excitation $W(t)$ local on site one, as in Eq.~(\ref{eq.hm.dimer}), we now investigate the effect of multiple excitations, choosing a periodic sequence, cf.~Fig.~\ref{fig.dimer2}a. 
Interestingly, we observe a successive increase of the total double occupation, $d=\sum_i\langle \chat{n}_{i\uparrow}\chat{n}_{i\downarrow}\rangle$, in the system until a value $d/L\approx1/4$ is reached. This final value is consistent with the time evolution of the many-particle energy spectrum
\begin{align}
\label{eq.hm.chain.spectrum}
S(\hbar\omega,t)=\sum_i |\langle\psi(t)|E_i\rangle|^2\textup{e}^{-\tfrac{(\hbar\omega-[E_i-UL/4])^2}{2(\hbar\omega_0)^2}}\,,
\end{align}
where $|E_i\rangle$ denote the energy eigenstates, which is shown in Fig.~\ref{fig.dimer2}c for a level broadening $\hbar\omega_0=J$. The final energy spectrum  ($t>200\hbar J^{-1}$) becomes symmetric around $\omega=0$, therefore providing on average two doublons in the system, corresponding to $d/L\rightarrow1/4$. Moreover, we observe that the double occupation (just as the density) becomes homogeneous along the chain, cf.~Fig.~\ref{fig.dimer2}b. At the same time, the correlation part of the interaction energy vanishes almost completely indicating approach of a mean-field state. In Fig.~\ref{fig.dimer2}d,
we furthermore investigate the same scenario for different fillings~\cite{rausch_17} ($N_\sigma=\sum_i n_{i\sigma}=1,2, \ldots,7$), which shows that the change of double occupation with respect to the initial ground state is largest for half-filling.

In order to test whether our doublon production protocol can be realized also in larger systems and 2D setups as well, we have performed extensive nonequilibrium Green functions simulations for $L$ up to $54$ of long duration, $t\le 400\hbar J^{-1}$. We used second-order Born self-energies within the generalized Kadanoff--Baym ansatz with Hartree--Fock propagators (HF-GKBA), as explained in detail in Refs.~\cite{balzer_lnp_13,schluenzen_cpp_16,schluenzen_prb_16, balzer_prb_16}. From benchmarks against density matrix renormalization group simulations~\cite{schluenzen_prb_17} we expect that these simulations are reliable and the results accurate, with an error of the double occupations $d$ not exceeding a few percent. Tests against  exact diagonalization simulations for the present excitation scenario confirm this result and show that the NEGF data for $d$ are approximately $10\%$ too low. 
\begin{figure}[t]
\includegraphics[width=0.48\textwidth]{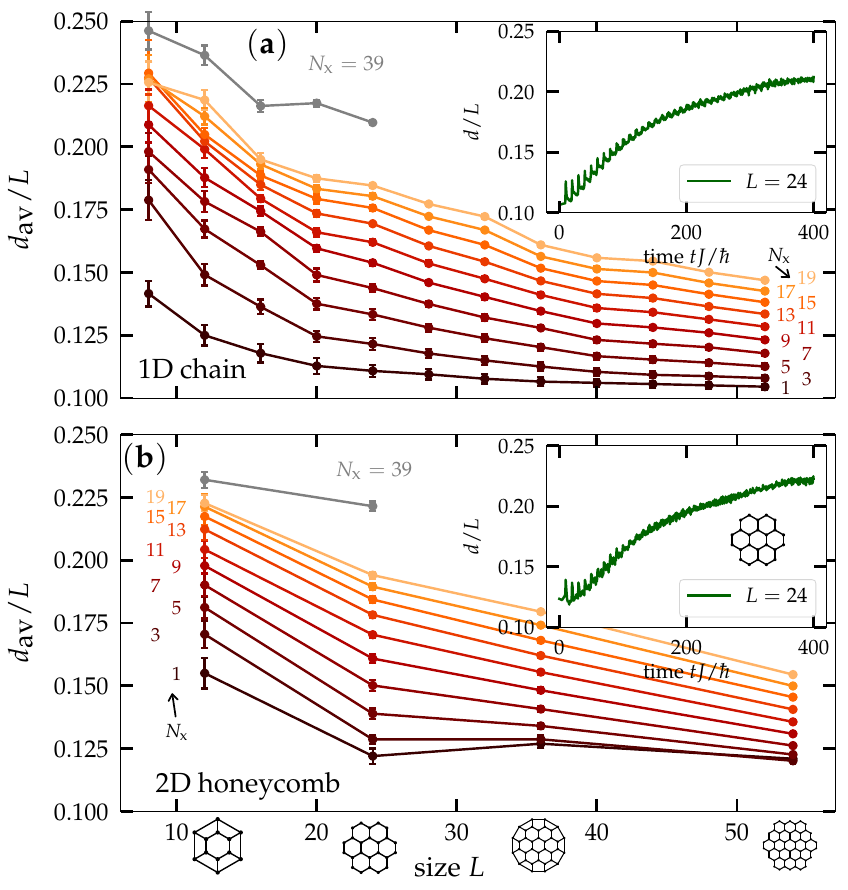}
\caption{\label{fig.GKBA} Extension of the periodic excitation protocol to larger systems of $L$ sites, using NEGF simulations (second Born self-energy with HF-GKBA~\cite{balzer_lnp_13}) for half-filled Hubbard clusters with $U=4J$. Mean final double occupation $d_\textup{av}/L$ for \textbf{a)}~1D chains and \textbf{b)}~2D honeycomb clusters of different size. The number of excitations, $N_\textup{x}$, is indicated in the figure. Insets show the time evolution of the mean double occupation as a function of time for $L=24$ and up to 40 excitations.}
\end{figure}

In Fig.~\ref{fig.GKBA} we show the final
average double occupation, $d_\textup{av}$, for a large range of system sizes $L$ after multiple ($N_\textup{x}=1\dots 39$) localized excitations of the same form as in Fig.~\ref{fig.dimer2}a, but with $W_0 = 2U = 8J$ for 1D chains, Fig.~\ref{fig.GKBA}a, and 2D honeycomb lattice fragments, Fig.~\ref{fig.GKBA}b. Clearly, the successive increase of $d_\textup{av}$ with $N_\textup{x}$ is confirmed for larger systems. Also, for fixed $N_\textup{x}$,  we observe a decrease of $d_\textup{av}$ with $L$, as expected. Extrapolating to larger values of $N_\textup{x}$ and taking into account the underestimation of $d_\textup{av}$ in our NEGF simulations (see above), we expect that an asymptotic value of $d_\textup{av}\to 0.25$ will be reached for all systems.

{\bf Summary and discussion}. In summary, we have presented a novel scenario for the production of doubly occupied electronic states in correlated finite 1D and 2D Hubbard clusters that is based on the impact of energetic ions. Our results are based on exact diagonalization simulations, for system sizes $L\le 12$, and nonequilibrium Green functions simulations, for $L\le 54$.
The physical mechanism has been made transparent by analytically solving the relevant dimer problem exposed to ion impact: it is the formation of avoided level crossings between bands of different doublon number, cf. Fig. \ref{fig.dimer2}, and it is straightforwardly extended to multiple sequential excitations.

For the case that the system is not coupled to a bath, which was analyzed in our simulations, the observed delocalized double occupation corresponds to a stationary nonequilibrium state which provides another example for pre-thermalization phenomena \cite{eckstein_prl_08,kollar_11,joura15,canovi_16} that recently have attracted high interest. More generally, we have presented a new scenario of nonequilibrium dynamics without thermalization \cite{rigol_nat_08, cramer_prl_08} that is driven by a localized single-particle potential quench instead of an interaction quench.
A suitable candidate to realize this scenario experimentally are fermionic atoms in optical lattices. Here the projectile dynamics could be accurately mimicked by a localized time-dependent variation of the lattice potentials of the sites adjacent to the impact point---a method that was demonstrated in Ref.~\cite{kuhr_nat_11}.

Another application of our results is correlated solid state systems, such as graphene fragments, that are exposed to energetic ions. 
For moderately correlated systems with typical parameters $J=1$eV and $a=1$\AA, the required
 ion velocities are of the order of $v_z=1aJ/\hbar$: for protons (alpha particles) this corresponds to a kinetic energy of $~120$\,eV ($480$\,eV). These are values that are well feasible with ion guns or in low-temperature high-pressure plasmas \cite{plasma-road-map-17} where the present effect has a strong influence on the stopping power \cite{puska14,zhao,balzer_prb_16} and may offer new applications. Of course, for the case of multiple excitations, one would need to consider spatial variations of the impact point, energy and time delay between impacts. These issues are easily studied within the dimer model and with our nonequilibrium Green functions approach as well. 
Furthermore, for these systems the coupling to the environment (bath) and the associated dissipation effects will have to be included. These effects will set an upper limit for the life time of the nonequilibrium doublon state in the range of several hundred femtoseconds. Since the time scale of the doublon formation is of the order of $1\ldots10$\,fs we expect that the presented scenario of multiple ion impacts can be realized.

\section*{Acknowledgements}
This work was supported 
by grant shp00015 for CPU time at 
the Norddeutscher Verbund f\"ur Hoch- und H\"ochstleistungsrechnen (HLRN).

\end{document}